\documentstyle[aps,preprint,epsf]{revtex}
\tighten

\newcommand{\pow}[3]{\left(\frac{#1}{#2}\right)^{#3}}
\def\rmd{{\rm d}}
\def\F{{\rm F}}
\def\SM{{\rm SM}}
\def\SUSY{{\rm SUSY}}

\begin{document}
\draft
\small
\preprint{OUTP-01-47-P}

\title{A C{\tt ++} Code to Solve the DGLAP Equations Applied to \\
  Ultra High Energy Cosmic Rays}
\author{\large Ramon Toldr\`a\thanks{toldra@thphys.ox.ac.uk}}
\address{Theoretical Physics, University of Oxford,
          1 Keble Road, Oxford OX1 3NP, UK}
         
\maketitle
\begin{abstract}
  We solve numerically the Dokshitzer-Gribov-Lipatov-Altarelli-Parisi
  (DGLAP) equations for the evolution of fragmentation functions using
  the Laguerre method. We extend this method to include supersymmetric
  evolution. The solution to the DGLAP equations is particularly
  interesting to calculate the expected spectra of Ultra High Energy
  Cosmic Rays in models where they are produced by the decay of a
  massive particle $X$, $M_X > 10^{12}$ GeV.
\end{abstract}
\vspace{1cm}

\newpage
{\large \bf PROGRAM SUMMARY}
\vspace{1cm}

{\it Title of program: } evolve

{\it Computer and operating system: } Program tested on Sun running
SunOS 5.7, Alpha running OSF1 4.0, Dell-PC running Linux Mandrake 7.2.

{\it Programming language used: } C{\tt++} with {\tt g++} compiler

{\it No. Lines in distributed program: } 2500


{\it Nature of Physical Problem: } In order to predict the spectra of
UHECR produced by the decay of a dark matter superheavy particle with
mass $M_X$ one needs to calculate fragmentation functions at the
energy scale $M_X$. These can be calculated from measured low energy
fragmentation functions using the DGLAP equations.

{\it Method of solution: } The DGLAP equations are solved by expanding
them in Laguerre polynomials which reduces their integration to the
computation of a set of coefficients. These coefficients are given by
algebraic recursive relations.

{\it Typical running time: } A few seconds.

{\it Restriction of the program: } Gluon coherence at $x<10^{-3}$ is
not included.

\newpage
{\large \bf LONG WRITE-UP}
\section{Introduction}

Over one hundred cosmic ray events with an energy higher than $4\times
10^{19}$ eV have been observed by different observatories (see
\cite{NaganoWatson} for a recent review). Were their sources at
cosmological distances, $d > 50-100$ Mpc, they would interact with the
Cosmic Microwave Background (CMB) on their way to the Earth and lose
their enormous energy. Therefore, one expects the sources for these
Ultra High Energy Cosmic Rays (UHECR) to be not far from the Galaxy.
However, there a few astrophysical sites in the galactic neighbourhood
which could accelerate a charged particle to such high energy. Faced
with this conundrum it has been suggested that UHECR are not
accelerated at all but are created at this ultra high energy by the
decay of massive dark matter particles $X$ generated in the early
universe \cite{KR,BKV,BS}. In order to test this hypothesis one needs
to calculate the angular distribution of events
\cite{EvansFerrerSarkar} and the spectra
\cite{Rubin,FodorKatz,BK,SarkarToldra} produced by the decay of the
population of $X$ particles clustered in the galactic halo.. In the
present work we will concentrate on the calculation of the spectra.

The $X$ decay contribution to UHECR is proportional to the inclusive
decay width of particle $X$, with mass~$M_X$, into particle~$h$
\cite{SarkarToldra}
\begin{equation} 
  \label{eq:DecayRate}
  \frac{1}{\Gamma_X}
  \frac{\rmd \Gamma(X \rightarrow h + \dots)}{\rmd x} = 
  \sum_a \int^1_x \frac{\rmd z}{z}\,
  \frac{1}{\Gamma_a}
  \left.\frac{\rmd \Gamma_a(y,\mu^2,M_X^2)}{\rmd y}\right|_{y=x/z} 
  D^h_a (z,\mu^2).
\end{equation}
Here $x$ is the fraction of the energy of $X$ carried by $h$ and $z$
is the fraction of the energy of parton~$a$ carried by $h$.  The first
factor in the integrand, the decay width of $X$ into parton~$a$,
$\rmd\Gamma_a/\rmd\,y$, is calculable in perturbation theory. It
encapsulates all kinematical effects in many-body decay
\cite{SarkarToldra}. In lowest order and for two-body decay it is
proportional to $\delta(1-y)$. The second factor, the non-perturbative
$D^h_a$, is the fragmentation function (FF) for particles of type $h$
from partons of type $a$. The mass scale $\mu$ is the factorization
scale, $\mu\sim M_X$~\cite{NasonWebber}. Particle $h$ is any final
state: $n$, $p$, $\gamma$, $e$, $\nu_e$, $\nu_\mu$ or $\nu_\tau$.

The FF satisfy a set of coupled integro--differential equations,
the Dokshitzer--\-Gribov--\-Lipatov--\-Altarelli--\-Parisi (DGLAP)
equations \cite{AltarelliParisi,DGL}. Given experimental data at some
low energy scale, say $M_Z$, an initial set of FF
$D^h_a(x,M_Z^2)$ can be extracted and evolved using the DGLAP
equations, to obtain the FF at some higher scale $D^h_a(x,M_X^2)$.

Hadronic structure functions (SF) satisfy as well DGLAP equations.
Although DGLAP equations for SF are similar to DGLAP equations for FF
they are not equivalent. Several approaches have been taken to solve
the DGLAP equations, usually applied to SF. One is the Mellin
transform method \cite{Mellin} which transforms the
integro--differential equations into ordinary differential equations.
However, at the end one needs to invert the Mellin transform to find
the solution in terms of $x$ which is a process with notorious
numerical problems. The QCDNUM program \cite{Botje} defines a grid in
$x$ and the energy scale $\mu^2$. The calculation of the SF on the
grid points is based on the computation of convolution integrals that
are evaluated as weighted sums.

A very elegant method to solve the DGLAP equations was introduced by
Furma\'nski and Petronzio~\cite{FurmanskiPetronzio}. It expands these
integro--differential equations using Laguerre polynomial so that
their integration is reduced to calculating a set of numerical
coefficients using simple algebraic recursive relations. The Laguerre
method has been implemented in numerical codes by several
groups~\cite{Ramsey,KumanoLondergan,CorianoSavkli}.

Usually numerical codes deal with SF, and all of them are oriented to
collider physics. Several include polarization and other aspects which
add complexity to the numerical evolution and that are irrelevant for
cosmic ray studies. Motivated as well by present or past experiments,
most of the numerical codes evolve SF using Standard Model equations,
without considering supersymmetry (SUSY) or any model beyond the
Standard Model (SM). The theoretical basis of SUSY DGLAP evolution for
SF was studied in Refs.~\cite{KounnasRoss,JonesLlewellyn}.  Numerical
solutions for the evolution of SF in several supersymmetric scenarios
have recently been presented in Ref.~\cite{Coriano}. In UHECR studies,
including SUSY is of paramount importance. Most of the evolution from
the low energy scale $M_Z$ to the high energy scale $M_X$ will be
governed by SUSY equations as long as SUSY is a symmetry of nature and
the SUSY breaking scale is of the order of the weak scale.

We have written a numerical code to evolve FF using the DGLAP
equations. It has been written bearing in mind its application to
UHECR and therefore does not include physics aspects such as spin
dependent functions that are relevant for collider physics but not
for UHECR physics. It includes SM evolution and SUSY
evolution. We have chosen the Laguerre method to solve numerically the
DGLAP equations. We have generalised the Laguerre method to include
SUSY evolution, which is not a trivial task, at least in the way that
this method is presented in the literature. 

Matrix algebra becomes very important when solving SUSY DGLAP
evolution by the Laguerre method. Therefore, it is convenient to code the
algorithm using an object oriented language. We have chosen C{\tt ++}, which
in the present context provides a good framework to perform matrix
calculations. The use of templates available in this programming
language simplifies the code. The final result is a fast algorithm
with good accuracy for the relevant range of $x$ and $\mu^2$.

\section{DGLAP Equations}

\subsection{General Equations, Leading Order and Notation}

The DGLAP equations can be written as
\begin{equation}
  \label{eq:AP}
  \frac{\partial D^h_a(x,\mu^2)}{\partial\ln\mu^2} = \sum_b
  \frac{\alpha_s(\mu^2)}{2\pi}P_{ba}(x,\alpha_s(\mu^2))\otimes D^h_b(x,\mu^2),
\end{equation}
where $\alpha_s(\mu^2)$ is the strong coupling constant and
$P_{ba}(x,\alpha_s)$ is the splitting function for the parton
branching $a\rightarrow b$. Here the convolution of two functions $A(x)$
and $B(x)$ is defined as
\begin{equation}
  \label{eq:Convolution}
  A(x)\otimes B(x) \equiv \int^1_x \frac{\rmd z}{z}\, A(z)B(\frac{x}{z}).
\end{equation}
The splitting functions can be expanded perturbatively
\begin{equation}
  \label{eq:LO}
  P_{ba}(x,\alpha_s)=P_{ba}(x)+O(\alpha_s)
\end{equation}
We will limit our study to leading order in $\alpha_s$ and therefore
ignore $O(\alpha_s)$ corrections to the splitting functions.

It is also convenient to define the following dimensionless evolution
parameter
\begin{equation}
  \label{eq:Tau}
  \tau \equiv \frac{1}{2\pi b} \ln \frac{\alpha_s(\mu^2_0)}{\alpha_s(\mu^2)},
\end{equation}
$b$ being the coefficient in the leading order beta function governing
the running of the strong coupling: $\beta(\alpha_s) = -b\alpha_s^2$.
We take $D^h_a$ to represent the sum of particle $h$ and, if
different, its antiparticle $\bar{h}$.

\subsection{Standard Model Equations}

The Standard Model DGLAP equations for the evolution of fragmentations
functions are well-known \cite{NasonWebber,ESW}. There are two parton
species: quarks~$q_k$, $k=1,\dots n_\F$ and gluons~$g$, with $n_\F$ the
total number of flavours. Conventionally, one defines the following linear
combinations (for ease of notation the superscript $h$ is omitted)
\begin{eqnarray}
  \label{eq:Combinations}
  D_{q^+_k} &\equiv& D_{q_k}+ D_{\bar{q}_k} \\
  D_q &\equiv& \sum_k D_{q^+_k}\\
  D_{q^-_k} &\equiv& D_{q_k}- D_{\bar{q}_k} \\
  D_{Q_k} &\equiv& D_{q^+_k}- \frac{1}{n_\F}D_q.
\end{eqnarray}
The non-singlet functions $D_{q^-_k}$ and $ D_{Q_k}$ obey the
equations 
\begin{eqnarray}
  \label{eq:NonSinglet1}
  \partial_\tau D_{q^-_k} &=& P_{qq} \otimes  D_{q^-_k} \\
  \label{eq:NonSinglet2}
  \partial_\tau D_{Q_k} &=& P_{qq} \otimes  D_{Q_k},
\end{eqnarray}
while the evolution of the singlet function $D_q$ is coupled to that
of the gluon function $D_g$ as
\begin{equation}
  \label{eq:SM2X2}
  \partial_\tau 
  \left(
    \begin{array}{l}
      D_q \\
      D_g
    \end{array}
  \right)
  =
  \left(
    \begin{array}{cc}
      P_{qq} & 2n_\F P_{gq} \\
      P_{qg} & P_{gg}
    \end{array}
  \right)
  \otimes
  \left(
    \begin{array}{l}
      D_q \\
      D_g
    \end{array}
  \right).
\end{equation}
The splitting functions were calculated in
Refs.~\cite{AltarelliParisi,ESW}.  Given the FF at some initial scale
$\mu_0$ for the quarks $q_k$ and gluon $g$,
Eqs.~(\ref{eq:NonSinglet1}--\ref{eq:SM2X2}) completely determine their
evolved values at some other scale $\mu$, to leading order in
$\alpha_s$.

\subsection{Supersymmetric Equations}

In a supersymmetric model (SUSY), besides the quarks and gluon one has the
superpartners: squarks~$s_k$ and gluinos~$\lambda$. In addition to the linear 
combinations~(\ref{eq:Combinations}) one now defines
\begin{eqnarray}
  \label{eq:SCombinations}
  D_{s^+_k} &\equiv& D_{s_k}+ D_{\bar{s}_k} \\
  D_s &\equiv& \sum_k D_{s^+_k}\\
  D_{s^-_k} &\equiv& D_{s_k}- D_{\bar{s}_k} \\
  D_{S_k} &\equiv& D_{s^+_k}- \frac{1}{n_\F}D_s.
\end{eqnarray}
The non-singlet function $ D_{q^-_k}$ and $ D_{s^-_k}$ evolve together 
as do $D_{Q_k}$ and $D_{S_k}$
\begin{eqnarray}
  \label{eq:SUSY2X2a}
  \partial_\tau 
  \left(
    \begin{array}{l}
      D_{q^-_k} \\
      D_{s^-_k}
    \end{array}
  \right)
  &=&
  \left(
    \begin{array}{cc}
      P_{qq} & P_{sq} \\
      P_{qs} & P_{ss}
    \end{array}
  \right)
  \otimes
  \left(
    \begin{array}{l}
      D_{q^-_k} \\
      D_{s^-_k}
    \end{array}
  \right) 
  \\
  \label{eq:SUSY2X2b}
  \partial_\tau 
  \left(
    \begin{array}{l}
      D_{Q_k} \\
      D_{S_k}
    \end{array}
  \right)
  &=&
  \left(
    \begin{array}{cc}
      P_{qq} & P_{sq} \\
      P_{qs} & P_{ss}
    \end{array}
  \right)
  \otimes
  \left(
    \begin{array}{l}
      D_{Q_k} \\
      D_{S_k}
    \end{array}
  \right).
\end{eqnarray}
The singlet functions for quarks and squarks, $D_q$ and $D_s$, are
coupled to the gluon and gluino functions, $D_g$ and $D_\lambda$, as 
\begin{equation}
  \label{eq:SUSY4X4}
 \partial_\tau 
  \left(
    \begin{array}{l}
      D_q \\
      D_g \\
      D_s \\
      D_\lambda
    \end{array}
  \right)
  =
  \left(
    \begin{array}{cccc}
      P_{qq} & 2n_\F P_{gq} & P_{sq} & 2n_\F P_{\lambda q} \\
      P_{qg} & P_{gg} & P_{sg} & P_{\lambda g}  \\
    P_{qs} & 2n_\F P_{gs} & P_{ss} & 2n_\F P_{\lambda s} \\
    P_{q\lambda} & P_{g\lambda} & P_{s\lambda} & P_{\lambda\lambda}
    \end{array}
  \right)
  \otimes
  \left(
    \begin{array}{l}
      D_q \\
      D_g \\
      D_s \\
      D_\lambda
    \end{array}
  \right). 
\end{equation}
In leading order the SUSY Eqs.~(\ref{eq:SUSY2X2a}--\ref{eq:SUSY4X4})
allow us to calculate the FF for all quark and squark flavours, gluons
and gluinos at some scale $\mu$ once their values at some initial
scale $\mu_0$ are known.  The SUSY DGLAP equations have been given in
the literature to leading order for structure functions
\cite{KounnasRoss,JonesLlewellyn}. Here we have presented their form
for FF. It is easy to see that one just needs to transpose the matrix
elements keeping the $n_\F$ factors in the same place to move from SF
to FF equations. The SUSY splitting functions were calculated in
Refs.~\cite{KounnasRoss,JonesLlewellyn}.

\section{Laguerre Algorithm}

\subsection{Evolution Operator}

In numerical studies it is better to consider the quantity
$xD(x,\tau)$. The evolution equations for $xD$ are the same as those
for $D$ if one multiplies the splitting functions by $x$. This
improves numerical stability since the $1/x$ singularity shown by
splitting functions with a gluon in the final state cancels off.
In general one has
\begin{equation}
  \label{eq:GeneralForm}
  \partial_\tau xD(x,\tau) = xP(x)\otimes xD(x,\tau),
\end{equation}
where $D$ is a $d$-dimension vector and $P$ is a $d\times d$ matrix
whose elements are splitting functions.
In the SM $d=1$ for Eqs.~(\ref{eq:NonSinglet1})
and~(\ref{eq:NonSinglet2}) while $d=2$ for Eq.~(\ref{eq:SM2X2}). In a
SUSY model $d=2$ for Eqs.~(\ref{eq:SUSY2X2a}) and~(\ref{eq:SUSY2X2b})
whereas $d=4$ for Eq.~(\ref{eq:SUSY4X4}).

Following Furma\'nski and Petronzio \cite{FurmanskiPetronzio} one
introduces the evolution operator $E(x,\tau)$
\begin{equation}
  \label{eq:EvolutionOp}
  xD(x,\tau)=E(x,\tau)\otimes xD(x,0).
\end{equation}
The evolution operator is a $d\times d$ matrix which satisfies the
following integro-differential equation
\begin{equation}
  \label{eq:DifferentialEq}
  \dot{E}(x,\tau)=xP(x)\otimes E(x,\tau)
\end{equation}
and the initial condition
\begin{equation}
  \label{eq:ICond}
  E(x,0)= \delta(1-x).
\end{equation}
One introduces the Laguerre expansions
\begin{eqnarray}
  \label{eq:LagExpP}
  e^{-x}P(e^{-x}) &=& \sum^\infty_{n=0} (xP)_n L_n(x) \\
  \label{eq:LagExpE}
  E(e^{-x}) &=& \sum^\infty_{n=0} E_n L_n(x),
\end{eqnarray}
with $L_n(x)$ the Laguerre polynomials of order $n$, which form an
orthonormal basis in the interval $(0,\infty)$ with weight $\exp -x$.
Equivalently, $L_n(-\ln x)$ are an orthonormal base in the interval
$(0,1)$ with unity weight. A key property of the Laguerre polynomials
is their closure under convolution
\begin{equation}
  \label{eq:Closure}
  \int^{x}_0 \rmd z\, L_n(z)L_m(x-z)=L_{n+m}(x)-L_{n+m+1}(x).
\end{equation}
Substituting Eqs.~(\ref{eq:LagExpP}) and~(\ref{eq:LagExpE}) into
Eq.~(\ref{eq:DifferentialEq}), using Eq.~(\ref{eq:Closure}) and the
fact that the Laguerre polynomials form a vectorial base one gets
the following system of linear equations
\begin{equation}
  \label{eq:EvolEqs}
  \dot{E}_n(\tau) = \sum^n_{m=0} (x\tilde{P})_{n-m} E_m(\tau),
\end{equation}
being
\begin{eqnarray}
  \label{eq:Ptilde}
  x\tilde{P}_0 &\equiv& xP_0, \\
  x\tilde{P}_m &\equiv& xP_m-xP_{m-1} \;\;\;\; m\ge 1.
\end{eqnarray}
Laguerre expansion of the initial condition Eq.~(\ref{eq:ICond}) 
translates into
\begin{equation}
  \label{eq:InitCond}
  E_n(\tau=0) = I.
\end{equation}
The Laguerre expansion transform an integro-differential equation into
a set of ordinary differential equations. Thus there is no need to
perform any intricate quadrature. One is left with an infinite
number of equations $n=0,1\dots \infty$. In practice one truncates the
Laguerre expansion at some finite $n=NLAST$.

\subsection{Expansion of the Splitting Functions}
\label{subsec:SplitFunc}

The coefficients of the Laguerre expansion of a function can be
quickly calculated if its Mellin transform is known. In the present
work we are interested in the product
\begin{equation}
  \label{eq:xP}
  F(x)\equiv xP(x),
\end{equation}
where $P(x)$ can be any splitting function. The Laguerre series is
\begin{equation}
  \label{eq:LaguerreExpand}
   F(e^{-y})=\sum^{\infty}_{n=0} F_n L_n(y).
\end{equation}
The Laguerre coefficients $F_n$ are given by
\begin{equation}
  \label{eq:LaguerreCoeff}
  F_n=\int^{\infty}_0 \rmd y\, e^{-y} L_n(y) F(e^{-y}),
\end{equation}
where $L_n(y)$ is the Laguerre polynomial of order $n$. Let $\hat{F}(s)$
be the Mellin transform of $F(x)$
\begin{equation}
  \label{eq:Mellin}
  \hat{F}(s)\equiv \int^1_0 \rmd x\, x^{s-1} F(x)
\end{equation}
then one can calculate the $F_n$'s by means of the formula
\cite{FurmanskiPetronzio}
\begin{equation}
  \label{eq:MellinLaguerre}
  \frac{1}{1-u}\hat{F}\left(\frac{1}{1-u}\right)=\sum^{\infty}_{n=0} F_n u^n.
\end{equation}
For the SM the Mellin transform of the splitting functions to leading
order have been listed in many works, see for example \cite{ESW}; for
a SUSY model the Mellin transforms of the splitting functions to
leading order are given in Ref.~\cite{KounnasRoss}. In both cases the
Mellin transforms of $P(x)$ are linear combinations of only six
simple functions: $1$, $1/(s+n)$ with $n=-1,0,1,2$ and
$\sigma(s)\equiv \gamma + \psi(s+1)$ where $\gamma$ is the Euler
constant and $\psi(s)$ is the digamma function. Using
Eq.~(\ref{eq:MellinLaguerre}) it is easy to show that each of these
six simple functions that appear in the Mellin transform
$\hat{(xP)}(s)=\hat{P}(s+1)$ gives a fixed contribution to $F_n$. The
rules to calculate the Laguerre coefficients of $xP(x)$ from the
Mellin transform of $P(x)$ are then:
\begin{eqnarray}
  \label{eq:Rules}
  \hat{P}(s) &\longrightarrow& \left(xP(x)\right)_n  \nonumber \\
  1 &\longrightarrow& 1 \\
  \frac{1}{s-1}  &\longrightarrow& \delta_{n0} \\
  \frac{1}{s}  &\longrightarrow&  \pow{1}{2}{n+1}\\
  \frac{1}{s+1}  &\longrightarrow& \frac{1}{2}\pow{2}{3}{n+1} \\
  \frac{1}{s+2}  &\longrightarrow& \frac{1}{3}\pow{3}{4}{n+1}\\
  \sigma(s) &\longrightarrow& 
  \left[\pow{1}{2}{n+1}+\delta_{n0}\right]-Z(n)(1-\delta_{n0}).
\end{eqnarray}
As usual one defines 
\begin{equation}
  \label{eq:Zn}
  Z(n)\equiv \sum^n_{m=1} (-1)^m\left( ^n_m \right)\zeta (m+1),
\end{equation}
where $\zeta(n)$ is the Riemann $\zeta$ function. All the Laguerre
coefficients for SM and SUSY splitting functions are listed in
Appendix~\ref{sec:LagSplitFunc}.

\subsection{Algorithm in $d$ dimensions} \label{sec:ddim}

Let us find the recursive relations that will allow us to solve the
DGLAP equations for any arbitrary dimension $d$. The quantities $E_n$
and $xP_n$ are $d\times d$ matrices. In analogy with the
$d=1,2$ formulae presented in~\cite{FurmanskiPetronzio}, let us try
the following ansatz for the solution to Eq.~(\ref{eq:EvolEqs}):
\begin{equation}
  \label{eq:Solution}
   E_n(\tau) = \sum^d_{i=1} e^{\lambda_i\tau}\sum^n_{k=0}
   \frac{\tau^k}{k!} B^k_{i,n},
\end{equation}
where the $\lambda_i$'s are the eigenvalues of $xP_0$. For every $i,k,n$
with $0\le k\le n$ $B^k_{i,n}$ is a constant $d\times d$ matrix; we
have in total $\frac{1}{2}(NLAST+1)(NLAST+2)\times d\times d^2$ matrix
elements. Substituting Eq.~(\ref{eq:Solution}) into
Eq.~(\ref{eq:EvolEqs}) we obtain the following two matrix relations 
for the $B^k_{i,n}$, $d>1$,
\begin{eqnarray}
  \label{eq:MatrixRel1}
  0 &=& \left(xP_0-\lambda_i I\right) B^n_{i,n}\\
  \label{eq:MatrixRel2}
  B^{k+1}_{i,n} &=& \left(xP_0-\lambda_i I\right)B^k_{i,n}+
  \sum^{n-1}_{m=k} x\tilde{P}_{n-m} B^k_{i,m},
\end{eqnarray}
while substitution of Eq.~(\ref{eq:Solution}) into
Eq.~(\ref{eq:InitCond}) gives
\begin{equation}
  \label{eq:MatrixRel3}
  I = \sum^d_{i=1} B^0_{i,n}.
\end{equation}
The matrix equation Eq.~(\ref{eq:MatrixRel1}) is equivalent to
$(NLAST+1)\times d \times d(d-1)$ algebraic equations, since $\det
\left(xP_0-\lambda_i I\right)=0$. Equation~(\ref{eq:MatrixRel2}) is
equivalent to $\frac{1}{2}NLAST(NLAST+1)\times d\times d^2$ equations
and Eq.~(\ref{eq:MatrixRel3}) is equivalent to $(NLAST+1)\times d^2$
equations. Adding up these three number we obtain
$\frac{1}{2}(NLAST+1)(NLAST+2)\times d\times d^2$ algebraic relations,
which is the same as the total number of matrix elements. Therefore,
the set of equations (\ref{eq:MatrixRel1}--\ref{eq:MatrixRel3})
completely determine all the matrix elements of $B^k_{i,n}$. They will
be given as recursive relations. In order to find them one defines the
projectors associated with the matrix $xP_0$
\begin{equation}
  \label{eq:Projectors}
  \Pi_i \equiv \prod_{j\neq i} \frac{xP_0-\lambda_j I}{\lambda_i-\lambda_j}.
\end{equation}
They have the following properties:
\begin{eqnarray}
  \label{eq:Prop1}
  \sum^d_{i=1} \Pi_i &=& I \\
  \label{eq:Prop2}
  \Pi_i\Pi_j &=& \Pi_i \delta_{ij} \\
  \label{eq:Prop3}
  \left(xP_0-\lambda_i I\right)\Pi_j &=& (\lambda_j-\lambda_i)\Pi_j
\end{eqnarray}
As the key step we define suitable auxiliary matrices for $d>1$:
\begin{eqnarray}
  \label{eq:bMatrices}
  b^0_{i,n}&\equiv& 0 \\
  b^k_{i,n}&\equiv& B^k_{i,n}-\sum_{j\neq i} (\lambda_j-\lambda_i)^k
  \Pi_j B^0_{i,n}, \;\;\; 0<k\le n.
\end{eqnarray}
Using Eqs.~(\ref{eq:MatrixRel1}--\ref{eq:MatrixRel3}) and the
projector properties Eqs.~(\ref{eq:Prop1}--\ref{eq:Prop3}) we have
found that all the $b^k_{i,n}$ and $B^k_{i,n}$ satisfy the following
relations:
\begin{eqnarray}
  \label{RecursiveEqs}
   B^0_{i,0} &=& \Pi_i \\
   b^0_{i,n} &=& 0 \\
   b^{k+1}_{i,n} &=& \left(xP_0-\lambda_i I\right)b^k_{i,n}+
   \sum^{n-1}_{m=k} x\tilde{P}_{n-m} B^k_{i,m} \\
   B^0_{i,n} &=& \Pi_i + \sum_{j\neq i} 
   \left( \frac{\Pi_i b^n_{j,n}}{(\lambda_i-\lambda_j)^n}-
     \frac{\Pi_j b^n_{i,n}}{(\lambda_j-\lambda_i)^n}\right) \\
   B^{k+1}_{i,n} &=& \left(xP_0-\lambda_i I\right)B^k_{i,n}+
   \sum^{n-1}_{m=k} x\tilde{P}_{n-m} B^k_{i,m}.
\end{eqnarray}
In the particular case $d=2$ these relations reduce to the ones given
in \cite{FurmanskiPetronzio}.  The above recursive relations can
easily be implemented in a computer using an object oriented language.
Then we can numerically calculate the $B^k_{i,n}$ once the matrices
$xP_n$ are given. The evolution operator is calculated using
Eq.~(\ref{eq:Solution}) without need to integrate numerically its
differential equation (\ref{eq:EvolEqs}) using any finite step
algorithm.

Finally, the evolved FF is given by the truncated series
\begin{eqnarray} \nonumber
  xD(x,\tau) &=& \sum^{NLAST}_{n=0}\sum^n_{m=0}
  E_{n-m}(\tau)(xD)_m\\ \label{eq:finalLag}
  & & \times \left(L_n(-\ln x)-L_{n+1}(-\ln x)\right),
\end{eqnarray}
where $(xD)_m$ are the coefficients in the Laguerre expansion
of the initial FF
\begin{equation}
  \label{eq:InitExp}
   xD(x,\tau=0) = \sum^\infty_{m=0} (xD)_m L_m(-\ln x).
\end{equation}

\section{Numerical Analysis}

\subsection{Evolution Steps}
For clarity we will assume flavour universality in the decay of $X$,
hence we will only consider the coupled singlet quark and gluon
evolution, Eq.~(\ref{eq:SM2X2}) for the Standard Model (SM), and
coupled singlet quark, gluon, singlet squark and gluino evolution,
Eq.~(\ref{eq:SUSY4X4}) for a SUSY model. Particular models for $X$ may
have different branching ratios for different flavours. The source
code includes all the routines necessary to evolve each quark and
squark flavour.

A (s)parton is not included in the evolution as long as the energy
scale is lower than its mass; when its threshold production scale is
crossed, it is added to the evolution equations with an initially
vanishing FF and it is assumed to be a relativistic particle.

In the SM case the code evolves the $q$ and $g$ initial FF from $M_Z$
to $M_t$, the top quark mass, with the number of flavours set to
$n_\F=5$, and then evolve from $M_t$ to $M_X$ with $n_\F=6$. Taking
$n_\F=6$ in the whole range from $M_Z$ to $M_X$ does not introduce any
significant difference in the final spectrum.

In the SUSY case the code evolves the $q$ and $g$ initial fragmentation
functions from $M_Z$ to the supersymmetry breaking scale $M_\SUSY >
M_t$ using the SM equations to obtain $D^h_i(x,M^2_\SUSY)$, with
$i=q,g$.  Then it takes $D^h_i(x,M^2_\SUSY)$, $i=q,g$, and
$D^h_j(x,M^2_\SUSY)=0$, $j=s,\lambda$, and evolves them from
$M_\SUSY$ to $M_X$ using the SUSY equations. All spartons are taken
to be degenerate with a common mass $M_\SUSY$. In the context of
structure functions a broken SUSY scenario with different masses for
$s$ and $\lambda$ was studied~\cite{Coriano}, showing no significant
difference with models with a unique SUSY mass. One expects the same
result to hold for FF.

In UHECR one is interested in the final spectra of baryons $p+n$,
neutrinos (sum of all three families) and photons.
The evolution of neutrinos and photons is equivalent, from a numerical 
point of view, to that of baryons. In the present work
we will concentrate on baryons. Initial FF are extracted from LEP data
at the energy scale $M_Z$ (see \cite{SarkarToldra} for further details
on initial FF and evolution of baryons, neutrinos and photons).

\subsection{Results}

\begin{figure}[tbh]
  \begin{center}
    \epsfxsize\hsize\epsffile{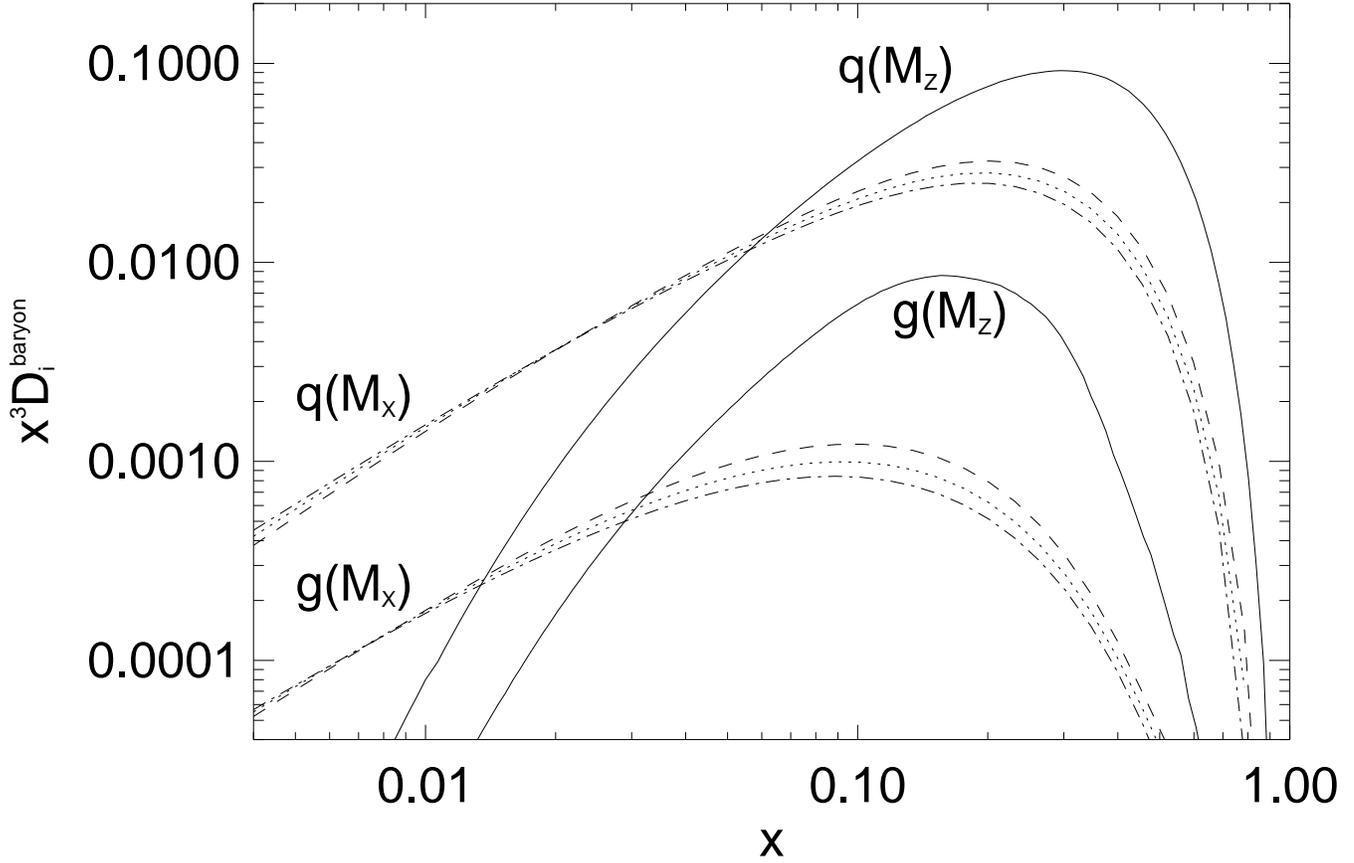}
    \bigskip
    \caption{Standard Model fragmentation functions 
      for baryons from quarks and gluons, at the initial scale
      $M_Z$ (solid lines) and the final scales: $M_X=10^{10}$~GeV (dashed
      line), $M_X=10^{12}$~GeV (dotted line) and $M_X=10^{14}$~GeV
      (dot-dashed line), showing scaling violations.}
    \label{fig:qgSM}
  \end{center}
\end{figure}

Let us first show SM evolution. In Fig.~\ref{fig:qgSM}
we plot the fragmentation functions for baryons from quarks and gluons
at the scale $M_Z$ (fit to LEP data) and their evolved shape at
$M_X=10^{10},10^{12}$ and $10^{14}$~GeV. Following the standard
convention in UHECR studies we always plot the quantity
$x^3D_a(x,\mu^2)$. One can see that as the final scale increases the
number of particles grows at low $x$ and diminishes at high $x$, a
well-known result from many previous studies of scaling violations.

\begin{figure}[tbh]
 \begin{center}
   \epsfxsize\hsize\epsffile{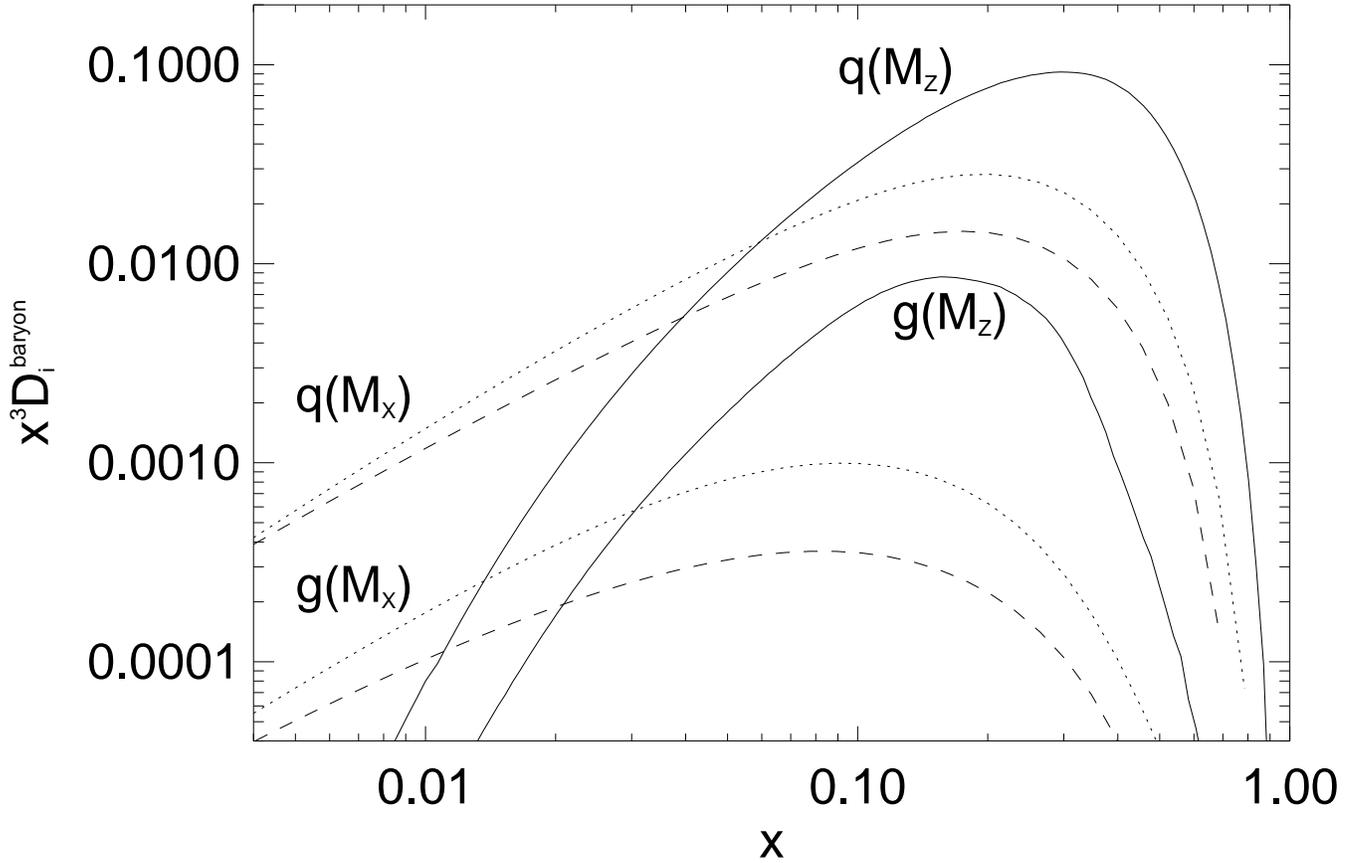}
   \bigskip
   \caption{Fragmentation functions for baryons from
     quarks and gluons, at the initial scale $M_Z$ (solid lines) and
     evolved to a decaying particle mass scale of $10^{12}$ GeV, for SM
     evolution (dotted lines), and, the more pronounced, SUSY evolution
     (dashed lines) taking $M_\SUSY=400$ GeV.}
  \label{fig:qgSM_SUSY} 
 \end{center}
\end{figure}

Next we compare SM evolution of FF with SUSY evolution. In
Fig.~\ref{fig:qgSM_SUSY} we show the common initial baryon curves at
$M_Z$, their shape after SM evolution up to $M_X=10^{12}$~GeV, and their
evolved shape at the same final scale after SUSY has been switched on
at $M_\SUSY=400$~GeV. It is clear that the SUSY curves have evolved
further than the SM curves. The difference between the two scenarios
stems chiefly from the different running of $\alpha_s(\mu^2)$. In a
SUSY model $\alpha_s$ decreases with the growth of energy scale more
slowly than in the SM because of the increased contribution to the
$\beta$-function from the SUSY partners. Since the rate of change
$\partial_{\ln \mu^2} D_a(x,\mu^2)$ is proportional to $\alpha_s$ (see
Eq.~(\ref{eq:AP})), a larger $\alpha_s$ translates into a larger
amount of evolution. In other words, for the same initial and
final scales we obtain $\tau_\SUSY (M_X)>\tau_\SM(M_X)$, using
Eq.~(\ref{eq:Tau}).

\begin{figure}[tbh]
 \begin{center}
   \epsfxsize\hsize\epsffile{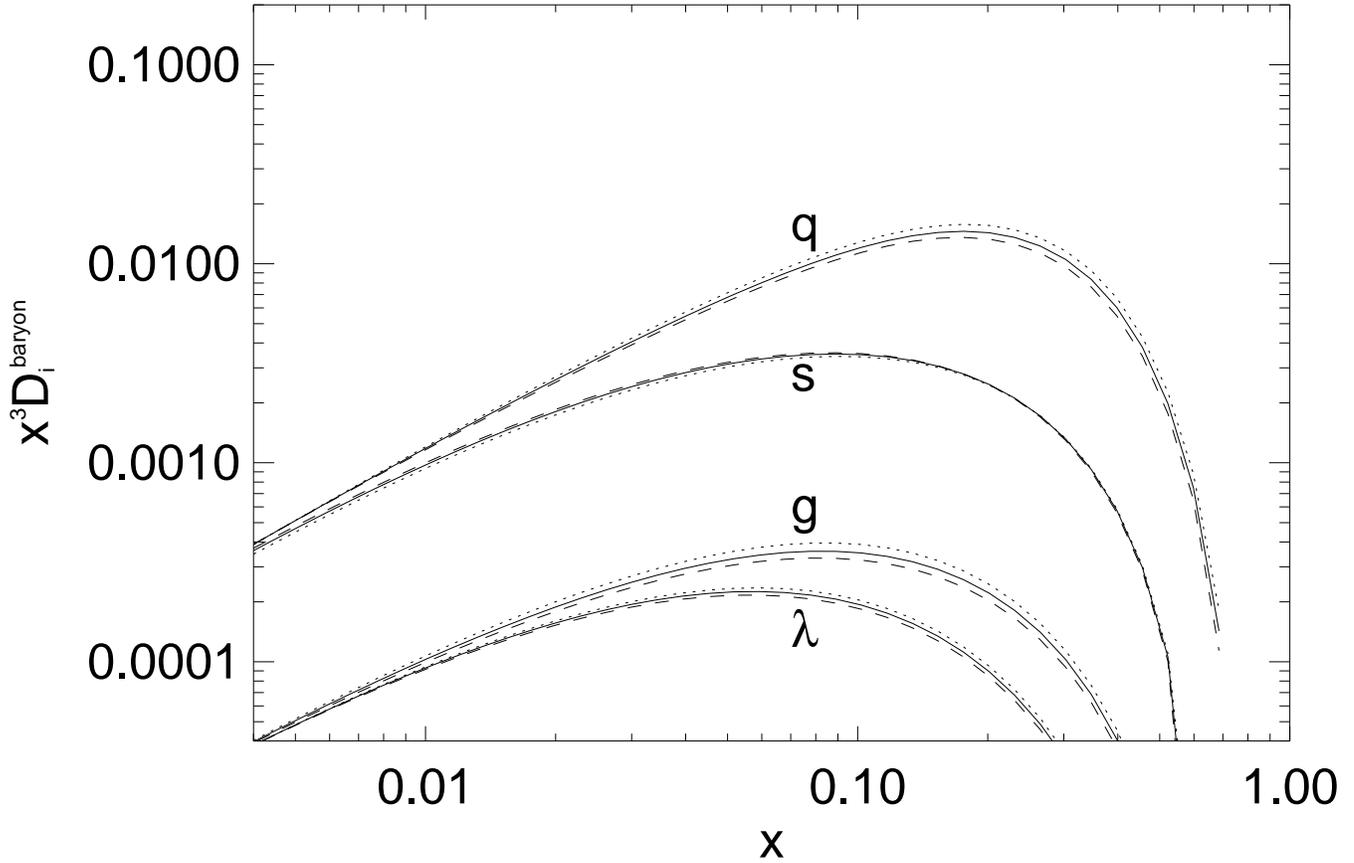}
   \bigskip
   \caption{Dependence on $M_{SUSY}$:
     the dashed lines are final (s)parton functions with
     $M_\SUSY=200$~GeV, the solid lines have $M_\SUSY=400$~GeV and the
     dotted lines have $M_\SUSY=1$~TeV. The initial and final scale are
     always $M_Z$ and $M_X=10^{12}$~GeV, respectively.}
   \label{fig:qgslSUSY} 
 \end{center}
\end{figure}

Figure~\ref{fig:qgslSUSY} shows the quark and gluon functions at
$M_Z$, their evolved values at $M_X=10^{12}$~GeV using the SUSY
equations for scales larger than $M_\SUSY$ and the radiatively
generated squark and gluino functions, all at the same final scale
$M_X$. We find that starting from vanishing values at $M_\SUSY$
the squark and gluino functions start to grow and catch up with the
quark and gluon functions, respectively, at small $x$ and from scales
a few orders of magnitude higher than $M_\SUSY$. This behaviour can be
understood qualitatively if one bears in mind that at low $x$ the
leading splitting function for quarks is
$2n_{\F}P_{gq}\sim4n_{\F}C_{\F}/x$, which is equal to the leading
splitting function for squarks $2n_{\F}P_{gs}\sim4n_{\F}
C_{\F}/x$. For gluons and gluinos the leading splitting functions tend
as well to a common value, $P_{gg}\sim2C_A/x$ and
$P_{g\lambda}\sim2C_A/x$, which is however different from that of
quarks and squarks.

SUSY evolution does not depend strongly on the chosen supersymmetry
breaking scale $M_\SUSY$. In Fig.~\ref{fig:qgslSUSY} we show the
curves obtained taking $M_\SUSY=200,400$~GeV and 1~TeV. The higher the
value of $M_\SUSY$, the less evolved the final curves for $q$ and
$g$. This follows from our comparison of the SM evolution and the SUSY
evolution. If SUSY switches on later (higher $M_\SUSY$) the energy
range over which the SM equations hold is larger. As we have seen
already, DGLAP evolution is slower when just the SM equations are
employed.

\subsection{Convergence, Accuracy and Scope} \label{sec:ConvAcc}

\begin{figure}[tbh]
  \begin{center}
    \epsfxsize\hsize\epsffile{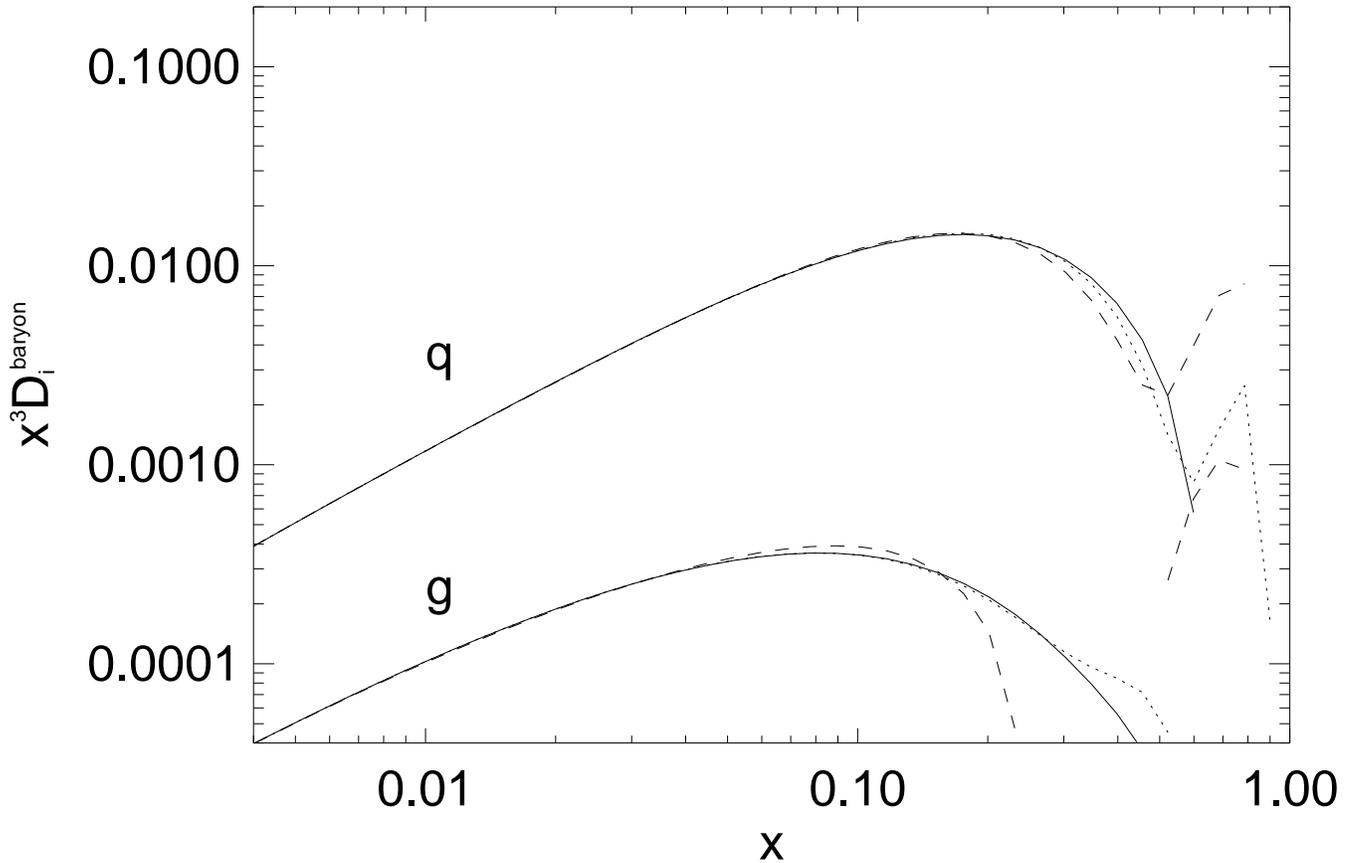}
    \bigskip
    \caption{SUSY evolution with $M_X=10^{12}$ GeV and
      $M_\SUSY=400$~GeV. The solid lines are obtained with $NLAST=15$,
      the dotted lines with $NLAST=12$ and the dashed lines with
      $NLAST=9$.  All curves have been calculated with {\tt alpha=0}.
      While good convergence is achieved very fast at intermediate
      values of $x$, for $x>0.2$ one needs to take a larger number of
      terms in the Laguerre expansion in order to damp out unphysical
      oscillations.}
    \label{fig:SUSYn}
  \end{center}
\end{figure}

The final FF generated by the Laguerre algorithm are accurate for
large values of the evolution parameter $\tau$.  For very small values
of $\tau$ the evolution operator approaches a delta function $\delta
(1-x)$ and thus truncation of the Laguerre series generates loss of
accuracy. The final result is stable for intermediate values of $NLAST$,
the values of $n$ at which the Laguerre expansion is truncated. In
Fig.~\ref{fig:SUSYn} we show SUSY evolution of $q$ and $g$ to the final scale
$M_X=10^{12}$ GeV, for $NLAST=9,12,15$. We see that the curves
approach to a common curve in most of the range of $x$. For large
values of $NLAST$ roundoff errors start to accumulate. In double 
precision, for SM evolution $NLAST$ has to be smaller than 30 while for
SUSY evolution, where the number of matrix operations
increases substantially, $NLAST$ has to be smaller than 20. 

In the limit $x\rightarrow 0$ the Laguerre polynomials with $n>0$
$L_n(-\ln x)$ go to infinity. One needs a large number of polynomials
to achieve good precision. However, taking $NLAST$ too large, roundoff
errors become important. The code give good accuracy for $x$ larger
than a few time $10^{-3}$. In any case, for $x<10^{-3}$ the DGLAP
equations (\ref{eq:AP}) no longer hold~\cite{ESW}. For very small $x$
gluon emission coherence has to be taken into account; this modifies the
kernel of the DGLAP equations. We have not included gluon coherence in
the present version of the code.

The functions $xD(x,\tau)$ fall rather fast when $x\rightarrow 1$.
This fact slows down the convergence of the Laguerre series in the
large $x$ region~\cite{FurmanskiPetronzio}. If the asymptotic
behaviour goes as $xD(x) \sim (1-x)^\alpha$ when $x\rightarrow 1$ then
one can improve the convergence by extracting the term $(1-x)^\alpha$
explicitly. The Laguerre series is then rewritten in terms of
generalised Laguerre polynomials $L_n^{(\alpha)}(-\ln
x)$~\cite{FurmanskiPetronzio}. Alternatively, one can take an
analytical approach and calculate the asymptotic exponent of
$xD(x,\tau)$ in the limit $x\rightarrow 1$~\cite{SarkarToldra}.
Nonetheless, one has to keep in mind that at large $x$ FF are not well
measured at low energy, hence, even in the case of perfect numerical
accuracy in the evolution, the final result would not be reliable for
$x > 0.6$.
   
\section{Description of the Input and Output} \label{sec:InOut}

To speed up the calculation the code takes as input the Laguerre
coefficients of the initial $xD$ at the scale $M_Z$. These are given
in the default files {\tt bar.ldat} for baryons, {\tt gam.ldat} for
photons and {\tt nu.ldat} for neutrinos. They have been obtained from
LEP data (baryons) or from the QCD Montecarlo generator
HERWIG~\cite{HERWIG} (photons and neutrinos), see
Ref.~\cite{SarkarToldra} for further details. These files contain two
fields: the Laguerre coefficients for the quark singlet function and
the Laguerre coefficients for the gluon function. The routine {\tt
  laguerre.cc} is provided in case one wishes to start with a
different set of initial data and needs to calculate the Laguerre
coefficients, see Sec.~\ref{sec:DescriptionCode}.

The main variables of the code are set with the help of a simple user
interface. There are two setup levels. The level 0 setup only allows
the user who runs the code to set the final scale in the evolution
$M_X$ which is stored in the variable {\tt double Efinal}. In this
case the evolution is supersymmetric with SUSY breaking mass
$M_\SUSY=400$ GeV and baryons are evolved. Level 1 gives more freedom
to the user. Besides the choice of $M_X$, now the user has to decide
between SM evolution or SUSY evolution (variable {\tt bool SUSY}), in
the case of SUSY evolution $M_\SUSY$ must be set (variable {\tt double
  SUSYMass}). The user has to select one particle to evolve: baryons,
photons or neutrinos. The variable {\tt string ParticleData} stores
the name of the file with initial data for the selected particle. The
standard template library class {\tt string} is included with the
header {\tt $<$string$>$}.

If $xD(x) \sim (1-x)^\alpha$, $\alpha >0$, when $x\rightarrow 1$,
the variable {\tt alpha} can be set to an integer larger than 0
to improve convergence for large $x$, as explained in Sec.~\ref{sec:ConvAcc}.
Level 0 defaults {\tt alpha=0}.

There are other variables that are initialised to default values,
which are the recommended values. To change them one needs to edit the
main program, recompile and link.  The variable {\tt double Einit}
stores the initial scale in the evolution. The default value is $M_Z$
but it could be set to any value with $M_b < {\tt Einit} < M_t$.
Remember to recalculate the initial Laguerre coefficients if {\tt
  Einit} is modified. The variable {\tt int NLAST} stores the number
of term in the Laguerre expansion ($n=0,1\dots ,NLAST$). The variables
{\tt xmin} and {\tt xmax} store the minimum and maximum values of $x$,
respectively, for the output of $xD(x)$ while {\tt NSTEPS} controls
the number of points in between.

The code evolves $xD$ up to the final scale $M_X$ and writes the final
functions to standard output. Output has three fields
for the SM ($x$, $xD_q$, $xD_g$) and five in a SUSY model ($x$, $xD_q$,
$xD_g$, $xD_s$, $xD_\lambda$).

\section{Description of the Code} \label{sec:DescriptionCode}

The following classes are defined in the code:
\begin{itemize} 
\item {\tt User}, {\tt User0}, {\tt User1}. They are defined in the
  files {\tt User.h} and {\tt User.cc}. They encapsulate the user
  interface that sets up the physical parameters of the run. The
  polymorphic class {\tt User} contains the virtual function {\tt
    setParameters()}, which prints a short help to standard error. The
  classes {\tt User0} and {\tt User1} inherit class {\tt User} and
  redefine {\tt setParameters()}. The function {\tt
    User0::setParameters()} manages the level 0 setup while {\tt
    User1::setParameters()} manages level 1.

\item {\tt StandardModel}, {\tt SUSYModel}. They are defined in the files {\tt
    className.h} and {\tt className.cc}. They encapsulate the physical
  model. Their member functions calculate the evolution parameter
  $\tau$ and the splitting functions $P_{ba}$. The class {\tt SUSYModel}
  inherits the class {\tt StandardModel} since at low energy any SUSY model
  must include the Standard Model. The class {\tt StandardModel} inherits
  class {\tt AuxFunc} (see next).
\item {\tt AuxFunc}, {\tt Laguerre}. They are defined in the files {\tt
    OtherClasses.h} and {\tt OtherClasses.cc}. The class {\tt AuxFunc}
  defines auxiliary functions required to calculate the Laguerre
  expansion of the splitting functions. In particular it stores
  tabulated values of the Riemann $\zeta$ function. The class {\tt
    Laguerre} encapsulates the Laguerre polynomials and the
  generalised Laguerre polynomials~\cite{FurmanskiPetronzio}.
\item {\tt Vector}, {\tt Matrix}. They are defined in {\tt Arrays.h} and {\tt
    Arrays.cc}. They are template classes with one template parameter
  which is the vector or square matrix size. The vector and matrix
  operators needed for the Laguerre algorithm are overloaded.
\item {\tt xD1X1}, {\tt xD2X2}, {\tt xD4X4}. They are defined in the
  header files {\tt className.h}. These classes perform the numerical
  integration by calculating the matrices $B$ using the recursive
  relations given in Sec.~\ref{sec:ddim}. The matrices $B$ and the
  evolution operator $E$ are members of these classes. Storage for the
  $B$ matrices is optimised taking into account that some indices are
  triangular, see Eq.~(\ref{eq:Solution}).  The class {\tt xD1X1}
  integrates the SM Eqs.~(\ref{eq:NonSinglet1})
  and~(\ref{eq:NonSinglet2}), {\tt xD2X2} integrates the SM
  Eq.~(\ref{eq:SM2X2}) and the SUSY Eqs.~(\ref{eq:SUSY2X2a})
  and~(\ref{eq:SUSY2X2b}), and {\tt xD4X4} integrates the SUSY
  Eq.~(\ref{eq:SUSY4X4}). All three classes are template classes with
  one generic data type which can be {\tt StandardModel} or {\tt
    SUSYModel}. All three classes inherit classes {\tt AuxFunc} and
  {\tt Laguerre}. In the present version of the code only the classes
  {\tt xD2X2$<$StandardModel$>$} and {\tt xD4X4$<$SUSYModel$>$} are
  used in the main program; the class {\tt xD1X1} is provided for
  completeness.
\end{itemize}

The main program {\tt Evolve.cc} takes the Laguerre coefficients of
the initial FF for $q$ and $g$ at $M_Z$ and calculates the final FF at 
$M_X$. It encodes the following steps:
\begin{enumerate}
\item It prompts the user for input and initialises variables.  It
  declares the input file stream {\tt qgFile} (class {\tt ifstream}
  included with {\tt $<$fstream$>$}) for the input quark singlet and
  gluon FF.
\item 
  It declares the objects {\tt sm} of type {\tt StandardModel} and
  {\tt susy} of type {\tt SUSYModel}. The public member functions {\tt
    StandardModel::setNFlavoursAndTau(EInit,EFinal)} and {\tt
    SUSYModel::setTau(susyMass,EFinal)} will be called to set the
  number of flavours (5 below $M_t$, 6 above) and calculate the
  evolution parameter $\tau$.
\item It declares as well the objects {\tt qg} of type
  {\tt xD2X2$<$StandardModel$>$} and {\tt qgsl} of type
  {\tt xD4X4$<$SUSYModel$>$}. The former solves the coupled evolution of
  singlet quark and gluon in the SM, while the latter solves the
  coupled evolution of singlet quark, gluon, singlet squark and gluino
  in a SUSY model. The public member functions
  {\tt xD2X2$<$StandardModel$>$::setB(sm)},
  {\tt xD2X2$<$StandardModel$>$::setE(sm)} calculate the matrices $B$
  and the evolution operator $E$ in the SM whereas
  {\tt xD4X4$<$SUSYModel$>$::setB(susy)},
  {\tt xD4X4$<$SUSYModel$>$::setE(susy)} calculate $B$ and $E$ in a
  SUSY model.
\item The member function {\tt setInitialxDLaguerre()} in class {\tt
    xD2X2$<$StandardModel$>$} is overloaded; it accepts an {\tt
    ifstream} type argument or {\tt void} argument. It reads the
  initial Laguerre expansion coefficients of $xD(x,\tau)$ or matches
  Laguerre coefficients at the $M_t$ scale. The mutator {\tt
    xD4X4$<$SUSYModel$>$::setInitialxDLaguerre(qg)} matches final SM
  Laguerre coefficients to initial SUSY Laguerre coefficients at
  $M_{SUSY}$.
\item The public member functions
  {\tt xD2X2$<$StandardModel$>$::setFinalxDLaguerre()} and
  {\tt xD4X4$<$SUSYModel$>$::setFinalxDLaguerre()} calculate the evolved
  Laguerre coefficients. The generalised Laguerre coefficients are
  calculated with the member functions {\tt setFinxDGenLag()}. Finally the
  accessor {\tt getxD(xl,xr,NSTEPS,alpha)} calculates the evolved
  fragmentation functions $xD(x,\tau)$ and writes the result to
  standard output. If {\tt alpha=0} the Laguerre
  coefficients are used, if {\tt alpha} is an integer larger than $0$ the
  generalised Laguerre coefficients are used instead.
\end{enumerate}

The header file {\tt CommonDefs.h} sets the precision to double,
defines some numerical factors and sets the maximum number of terms in
the Laguerre expansion {\tt NMAX=30}. The header file {\tt PhysicalConst.h}
defines some physical constants used in the computation.

As mentioned in Sec.~\ref{sec:InOut} default files with Laguerre
coefficients for the initial FF are provided. If one wishes to use
another set of initial data, one will need to calculate their Laguerre
coefficients. The file {\tt laguerre.cc} provides a routine to
calculate these coefficients when one has $xD$ in tabulated form, i.e.
one has a file with two fields: the firsts one is $x$ and the second
one $xD(x)$. The routine in {\tt laguerre.cc} declares an object of
class {\tt Laguerre} and therefore must be compiled with the file {\tt
  OtherClasses.cc} (the file {\tt Makefile} is provided with the
source files to facilitate compilation). Notice that the routine in
{\tt laguerre.cc} is independent of the main program in {\tt
  Evolve.cc}. The Laguerre coefficients in Eq.~(\ref{eq:InitExp}) are
given by
\begin{equation}
  \label{eq:LagCoeff}
  (xD)_n = \int^\infty_0 \rmd y\; e^{-y} (xD)(e^{-y}) L_n(y) =
  \int^1_0 \rmd x\; xD(x) L_n(-\ln x).
\end{equation}
The integral is performed numerically using the extended trapezoidal
rule. Previous implementations of the Laguerre method tend to use
parametric fits to the data of the sort $Ax^\alpha(1-x)^\beta$ at the
initial scale. In this case the Laguerre coefficients can be
calculated analytically and are given in terms of infinite sums.
However, the $x$ range of validity of this fits tends to be rather
narrow. Furthermore, initial data is naturally obtained in tabulated
form from experiments and from Montecarlo generators. Therefore we
prefer calculating the initial Laguerre coefficients as explained
above without use of any intermediate parametric fit in $x$ space.

\section{Conclusions}

In order to test models where UHECR are produced by the decay of
superheavy dark matter of mass $M_X$ one needs to calculate the
predicted spectra for baryons, photons and neutrinos. The $X$ produced
injection spectrum for any primary cosmic ray is basically given by
fragmentation functions at the energy scale $M_X$. We calculate FF by
evolving low energy FF up to the ultra high energy $M_X$ using the
DGLAP equations. We have solved numerically the DGLAP equations using
the Laguerre method. In our study we have considered two scenarios:
Standard Model evolution and SUSY evolution. We have generalised the
Laguerre method to include supersymmetry. The final result is a
numerical code which is fast and accurate for $2\times 10^{-3} <
x < 0.6$ and $M_X > 1000$ GeV. 

\section*{Acknowledgements}

I would like to thank Subir Sarkar for useful discussion and
encouragement. I acknowledge the support of the Marie Curie
Fellowship No.~HPMF-CT-1999-00268.

\appendix

\section{Laguerre Coefficients for the Splitting Functions}
\label{sec:LagSplitFunc}

The Laguerre expansions of the splitting functions can be calculated
using Eq.~(\ref{eq:MellinLaguerre}).
Let us first list the Laguerre coefficients of $xP_{ab}(x)$ in the SM
\cite{FurmanskiPetronzio,KumanoLondergan}:
\begin{eqnarray}
  \label{eq:SMSplitLaguerre}
  \left(xP_{qq}\right)_n &=& -\frac{4}{3} C_F \delta_{n0} + 2C_F\left( 
    Z(n)+\frac{1}{4}\left(3-\pow{1}{2}{n}-
      \pow{2}{3}{n+1}\right)\right)(1-\delta_{n0}) \\
  \left(xP_{gq}\right)_n &=& \frac{4}{3} C_F \delta_{n0} + 2C_F\left( 
    \frac{1}{4}\pow{2}{3}{n+1}-\pow{1}{2}{n+1}\right) (1-\delta_{n0}) \\
  \left(xP_{qg}\right)_n &=& T_R \left( \pow{1}{2}{n+1}-
    \pow{2}{3}{n+1}+\frac{2}{3}\pow{3}{4}{n+1}\right) \\
  \left(xP_{gg}\right)_n &=& -\frac{2}{3}T_R n_\F \delta_{n0} + \\ \nonumber
  && 2C_A\left(\left(Z(n)-\pow{1}{2}{n}+
      \frac{1}{2}\pow{2}{3}{n+1}-\frac{1}{3}\pow{3}{4}{n+1}+
    \frac{11}{12}\right)-\frac{2}{3}T_R N_F\right)(1-\delta_{n0})
\end{eqnarray}

Next, we present the Laguerre coefficients of $xP_{ab}(x)$ in a SUSY
model. To calculate them we use the table of Mellin transform given in
\cite{KounnasRoss} (in this reference there is a mismatch between the
normalization of the splitting functions and the normalization of the
DGLAP equations that we correct). From the rules given in
Subsec.~\ref{subsec:SplitFunc} we obtain
\begin{eqnarray}
  \label{eq:SUSYSplitLaguerre}
  \left(xP_{qq}\right)_n &=& -\frac{11}{16} C_F \delta_{n0} + 2C_F\left[ 
    Z(n)+\frac{1}{4}\left(2-\pow{1}{2}{n}-
      \pow{2}{3}{n+1}\right)\right](1-\delta_{n0}) \\
  \left(xP_{gq}\right)_n &=& \frac{4}{3} C_F \delta_{n0} + 2C_F\left[ 
    \frac{1}{4}\pow{2}{3}{n+1}-\pow{1}{2}{n+1}\right] (1-\delta_{n0}) \\
  \left(xP_{qg}\right)_n &=& T_R \left[ \pow{1}{2}{n+1}-
    \pow{2}{3}{n+1}+\frac{2}{3}\pow{3}{4}{n+1}\right] \\
  \left(xP_{gg}\right)_n &=& -\left(\frac{1}{3}C_A+T_R n_\F \right)
  \delta_{n0}+ \\
  \nonumber && \left[2C_A\left(Z(n)-\pow{1}{2}{n}+
      \frac{1}{2}\pow{2}{3}{n+1}-\frac{1}{3}\pow{3}{4}{n+1}+
      \frac{3}{4}\right)-T_R N_F\right](1-\delta_{n0}) \\
  \left(xP_{sq}\right)_n &=& C_F\frac{1}{3} \pow{2}{3}{n} \\
  \left(xP_{\lambda q}\right)_n &=& C_F \left[\pow{1}{2}{n+1}-
    \frac{1}{3}\pow{2}{3}{n}\right] \\
  \left(xP_{sg}\right)_n &=& T_R \left[\pow{2}{3}{n+1}-
    \frac{1}{2}\pow{3}{4}{n}\right] \\
  \left(xP_{\lambda g}\right)_n &=& C_A \left[\pow{1}{2}{n+1}-
    \pow{2}{3}{n+1}+\frac{1}{2}\pow{3}{4}{n}\right] \\
  \left(xP_{qs}\right)_n &=& C_F\pow{1}{2}{n+1} \\
  \left(xP_{gs}\right)_n &=& C_F\delta_{n0}-C_F\pow{1}{2}{n}(1-\delta_{n0}) \\
  \left(xP_{ss}\right)_n &=& -2C_F\delta_{n0}+C_F\left[2Z(n)-
    \pow{1}{2}{n} +1\right](1-\delta_{n0}) \\
  \left(xP_{\lambda s}\right)_n &=& C_F\pow{1}{2}{n+1} \\
  \left(xP_{q\lambda}\right)_n &=& T_R \left[\pow{1}{2}{n+1}-
    \frac{1}{3}\pow{2}{3}{n}\right] \\
  \left(xP_{g\lambda}\right)_n &=& \frac{4}{3}C_A \delta_{n0}+ 
  C_A \left[\frac{1}{3}\pow{2}{3}{n}-\pow{1}{2}{n}\right](1-\delta_{n0}) \\
  \left(xP_{s \lambda}\right)_n &=& T_R\frac{1}{3}\pow{2}{3}{n} \\
  \left(xP_{\lambda\lambda}\right)_n &=& -\left(\frac{4}{3}C_A+T_R n_\F \right)
  \delta_{n0}+ \\
  \nonumber && \left[C_A\left(2Z(n)-\pow{1}{2}{n+1}-\frac{1}{3}\pow{2}{3}{n}+
      \frac{3}{2}\right)-T_R N_F\right](1-\delta_{n0})
\end{eqnarray}

\newpage
{\large \bf TEST RUN INPUT AND OUTPUT}
\addtolength{\baselineskip}{-1mm}
\begin{verbatim}
% evolve

  Evolve fragmentations functions up to the energy scale M_X
  Two setup levels: 
  Level 0 -> Set M_X
  Level 1 -> Set M_X, Standard Model or SUSY evolution,
             M_SUSY (in the latter case), particle (baryon, 
             photon or neutrino), and alpha

Enter level (0 or 1): 1
Enter final scale M_X in GeV: 1e12
Enter 0 for Standard Model evolution, 1 for SUSY evolution: 1
Enter SUSY breaking scale M_SUSY in GeV: 400
Enter particle (b baryons, g photons, n neutrinos): n
Enter alpha, alpha >= 0 (type 0 if you don't know): 4

Fragmentations functions evolved up to M_X=1e+12 GeV
SUSY evolution with M_SUSY=400 GeV
Initial Laguerre coefficients read from file: ./init_data/nu.ldat
Parameter alpha=4

0.001 413.595 40.1029 360.98 37.8363 
0.00114574 372.335 35.3804 319.938 33.179 
0.00131271 334.828 31.14 282.839 29.0114 
0.00150402 300.755 27.3395 249.372 25.291 
0.00172321 269.822 23.9402 219.247 21.9782 
0.00197435 241.756 20.9062 192.194 19.0364 
0.00226209 216.308 18.2042 167.958 16.4314 
0.00259176 193.247 15.8036 146.305 14.1317 
0.00296947 172.365 13.6762 127.012 12.1079 
0.00340223 153.468 11.7957 109.873 10.3329 
0.00389806 136.381 10.1384 94.6951 8.78167 
0.00446615 120.945 8.68196 81.2995 7.4311 
0.00511703 107.013 7.40617 69.5181 6.25986 
0.00586277 94.454 6.29235 59.1949 5.24839 
0.00671719 83.1462 5.32339 50.1848 4.37872 
0.00769614 72.9806 4.48362 42.3533 3.63442 
0.00881775 63.8575 3.75871 35.5756 3.0005 
0.0101028 55.686 3.13559 29.7366 2.46334 
0.0115752 48.3833 2.60235 24.7301 2.0106 
0.0132621 41.8736 2.14816 20.4588 1.63113 
0.0151949 36.0872 1.76324 16.8335 1.31494 
0.0174093 30.9602 1.43872 13.773 1.05307 
0.0199465 26.4336 1.16663 11.2037 0.837566 
0.0228534 22.4526 0.939831 9.05914 0.661396 
0.026184 18.9664 0.75193 7.27963 0.518366 
0.03 15.9278 0.597254 5.812 0.40307 
0.0343721 13.2927 0.470784 4.60909 0.310818 
0.0393814 11.0197 0.368103 3.62942 0.237572 
0.0451207 9.07067 0.285354 2.83672 0.17988 
0.0516964 7.40964 0.219183 2.19954 0.134819 
0.0592305 6.00339 0.166701 1.69086 0.0999292 
0.0678626 4.82113 0.125434 1.28761 0.0731663 
0.0777527 3.83453 0.0932815 0.970304 0.052841 
0.0890841 3.01766 0.0684755 0.722594 0.0375722 
0.102067 2.34699 0.0495397 0.530891 0.0262409 
0.116942 1.80132 0.0352533 0.383979 0.0179478 
0.133985 1.36174 0.0246149 0.272665 0.0119761 
0.153511 1.01153 0.0168107 0.189457 0.00775903 
0.175883 0.736047 0.0111848 0.128271 0.00485098 
0.201516 0.522559 0.00721274 0.0841796 0.00290404 
0.230884 0.360077 0.00447843 0.0531969 0.0016483 
0.264532 0.239156 0.00265422 0.0321031 0.000876103 
0.303084 0.151691 0.00148427 0.0183037 0.000429735 
0.347255 0.0907179 0.000770882 0.00972273 0.000191815 
0.397863 0.0502393 0.000363635 0.00472289 7.78663e-05 
0.455846 0.0250863 0.00015074 0.00204541 3.02731e-05 
0.52228 0.0108333 5.21486e-05 0.000762316 1.30163e-05 
0.598395 0.00376894 1.37897e-05 0.000232293 6.639e-06 
0.685603 0.000921878 2.35586e-06 5.33011e-05 3.14451e-06 
0.785521 0.000115102 1.76694e-07 7.61964e-06 8.88797e-07 
0.9 2.33062e-06 1.74687e-09 2.87264e-07 5.27823e-08 
\end{verbatim}
\addtolength{\baselineskip}{+1mm}

\newpage

\end{document}